\documentstyle[11pt, titlepage]{article}

\newcommand{\beq}{\begin{equation}}
\newcommand{\beqn}{\begin{eqnarray}}
\newcommand{\eeq}{\end{equation}}
\newcommand{\eeqn}{\end{eqnarray}}
\begin{document}
\begin{titlepage}
\rightline {HUTP-95/A027}
\rightline {DART-HEP-95/04}
\rightline {astro-ph/9507108}
\begin{center}
\bigskip \bigskip \bigskip \bigskip \bigskip
\Large\bf Post-Inflation Reheating \\
in an Expanding Universe \\
\bigskip\bigskip\bigskip
\normalsize\rm
David I. Kaiser \\
\bigskip \it Lyman Laboratory of Physics\\
Harvard University\\
Cambridge, MA 02138\\ \rm
e-mail:  dkaiser@fas.harvard.edu\\
\bigskip\bigskip\bigskip\bigskip
31 October 1995\\ \bigskip
[Revised Draft] \\
\bigskip\bigskip\bigskip\bigskip
\end{center} \narrower
  An analytic means of studying the resonant
decay of the inflaton field is developed for the case of
background expansion, $H \neq 0$.  It is shown that the parametric
resonance in the inflaton's decay need not disappear when the
expansion of the universe is taken into account, although the total
number of particles produced is fewer than in the $H \simeq 0$
case.
\bigskip\\

PACS number:  98.80Cq \\
\end{titlepage}
\newpage

\baselineskip 24pt
\section{Introduction}
\indent  Soon after the first models of inflation were published in
1981-1982~\cite{alan}~\cite{newi}, several papers
were written to study how the universe would reheat
following the supercooled inflationary state.~\cite{reh}
The process appeared to be straightforward:  in models
like new inflation (and chaotic inflation~\cite{chaotic}) which
incorporate a second order phase transition to end
inflation, the inflaton field would wind up oscillating
around the minimum of its potential near the end of
inflation.  These oscillations would produce a sea of
relativistic particles, if one added (by hand) interaction
terms between the inflaton and these lighter species.  The decay rates
$\Gamma_{\phi \rightarrow \chi , \psi}$ could then be calculated, where
$\phi$
is the (decaying) inflaton field, $\chi$ is some light boson field, and
$\psi$ is some light fermion field.  From these decay rates, the energy
density ($\rho_p$) of the produced particles could be calculated, and
related to the
final reheat temperature ($T_{\rm rh}$):  $\rho_p \simeq \pi^2  N_{\rm
eff} \>  T_{\rm
rh}^4 / 30 $, where $N_{\rm eff}$ counts the effective number of massless
spin degrees of freedom.  This method was thought to estimate $T_{\rm
rh}$ to
within an order of magnitude. \\
\indent  Recently, much attention has been focused on an overlooked but
dramatic feature of the production-via-oscillation model:  the inflaton's
oscillations should be {\it unstable}, and should exhibit {\it parametric
resonance}.  Certain modes $\chi_k$ within small bands $\epsilon \ll k$
should grow {\it exponentially}, swamping the production in other modes.
This effect has been studied by Kofman, Linde,
and Starobinsky in~\cite{linde} for both the broad and narrow resonance
regimes, and, following different methods, by Shtanov, Traschen, and
Brandenberger~\cite{STB} for the narrow resonance regime.  The importance
of \lq\lq induced amplification" in the inflaton's decay has been
addressed recently with still different methods by Boyanovsky {\it et
al.} in~\cite{DB}~\cite{DB2}.  See also~\cite{dolfre}~\cite{Yosh}.
Earlier work on some of these questions may be found
in~\cite{TB90}~\cite{dolkir}.  The crucial feature of the new reheating
scenario is that the equation of motion for the light boson field will
not obey the simple harmonic oscillator equation.
Instead, for an interaction of the form ${\cal L}^{\prime}
= - \frac{1}{2} g^2 \phi^2 \chi^2$, the mode functions of the boson field
will obey:
\beq
\ddot{\chi}_k + 3H\dot{\chi}_k + \left( \frac{k^2}{a^2} +
g^2 \phi^2 \right) \chi_k = 0 ,
\label{eomH}
\eeq
where overdots denote derivatives with respect to cosmic time $t$,
$a(t)$ is the scale factor for the (assumed) flat
Friedmann-Robertson-Walker spacetime, and $H \equiv \dot{a} / a$ is the
Hubble parameter.  Since the inflaton field ($\phi (t)$) is rapidly
oscillating, for
frequencies $\omega_I \gg H$ the $3H \dot{\chi}_k$ term may be
neglected and $a(t)$ treated as constant.
Then this equation for the boson field mode functions will look
like the well-known Mathieu equation (see, {\it
e.g.},~\cite{handb}~\cite{L2}):
\beq
\ddot{\chi}_k + \omega_k^2 \left( 1 + h \cos \omega_I t
\right) \chi_k = 0 .
\label{Mathieu}
\eeq
As demonstrated in~\cite{linde}~\cite{STB}~\cite{TB90} and treated below,
for $\omega_I = 2
\omega_k + \epsilon$, with $\epsilon \ll \omega_k$, solutions grow like
$\chi_k (t)
\propto \exp (\pm \mu_{\pm} t)$, instead of like $\chi_k (t) \propto \cos
(\omega_k t)$.  This means that the number of $\chi$
bosons produced goes like $N_{\chi}^{\rm res} \propto \exp (2
\mu_{+} t)$, so that the decay rate for $\phi \rightarrow 2
\chi$ is much larger than originally calculated:
$\Gamma^{\rm res}_{\chi} \gg
\Gamma_{\chi}$.  (Note that this parametric
resonance is only effective for inflaton decay into bosons;
the exponential increase in fermion modes is forbidden
because of Fermi-Dirac statistics.)  This large decay rate increases
$\rho_p$, ultimately giving $T_{\rm rh}^{\rm res} \gg T_{\rm rh}$. \\
\indent  In this paper, we develop an analytic means of studying
the inflaton's decay into other particles for the case of $H \neq 0$.
We find that the total number of particles produced from the parametric
resonance effect when $H \neq 0$ is fewer than in the $H \simeq 0$ case,
but can still be exponentially greater than when the resonance is
neglected altogether.  Rather than expansion of
the universe, then, the true threat to resonant inflaton decays appears
to be
back-reaction of the produced particles on the amplitude of the decaying
inflaton field.  While we do not study this back-reaction effect in
detail here, some approximate schemes for including some of its effects are
discussed in the Appendix. \\
\indent  We concentrate on the case of the inflaton field
$\phi$ decaying into further inflaton bosons, via the self-interaction
potential $V(\phi)$.  We restrict attention to the
small-amplitude, narrow-resonance regime, as the inflaton field oscillates
near the minimum of
its potential, similar to that studied in~\cite{STB}.  In section 2,
the number of particles produced
per mode $k$, $N_k$, is calculated for non-zero expansion, assuming
that the decay is not resonant. Section 3 presents the calculation of
$N^{\rm res}_k$ when the expansion
of the universe is neglected.  In section 4 we calculate
$N^{\rm res}_k$ taking into account the expansion of the universe, and
demonstrate that the
parametric resonance need not vanish.  These values of $N_k$ correspond
to particle production during the narrow-resonance regime of the \lq\lq
preheating" epoch discussed in~\cite{linde}~\cite{preheat}.  Final reheating
temperatures
can then be calculated, as discussed in~\cite{linde}, using the methods
developed in~\cite{reh}, where these methods should now be applied to
the produced particles of the preheating epoch; this will not be included
here.  For more on the question of thermalization and the final reheat
temperature, see also~\cite{DB2}.  Concluding remarks follow in section 5.

\section{Non-Resonant Inflaton Decay in an Expanding Background}
\indent  We will study a very simple model of inflation:  chaotic
inflation with the potential $V(\phi) = \frac{1}{4} \lambda \phi^4$.  We
will assume that the metric may be written in the form of a flat
Friedmann-Robertson-Walker line element, $ds^2 = g_{\mu\nu} dx^{\mu}
dx^{\nu} = - dt^2 + a^2 (t) d \vec{x}^2$.  Then, given the
lagrangian density
\beq
{\cal L} = \sqrt{-g} \left[ \frac{1}{16\pi G} R - \frac{1}{2} \phi_{; \>
\mu} \phi^{; \> \mu} - V(\phi) \right] ,
\label{lag}
\eeq
the equations of motion take the familiar form:
\beqn
\nonumber H^2 &=& \frac{8\pi G}{3} \left[ V(\phi) + \frac{1}{2}
\dot{\phi}^2 \right] , \\
\ddot{\phi} &+& 3H \dot{\phi} - \frac{1}{a^2}\nabla^2 \phi + \frac{d V}{d
\phi} = 0 .
\label{eom}
\eeqn
Next we decompose the inflaton field into the sum of a classical background
field and a quantum fluctuation:  $\phi (\vec{x} , t) = \varphi (t) +
\delta \phi (\vec{x} , t)$.  (We will neglect the metric fluctuations
which couple to these inflaton fluctuations for the present analysis; for
more on these metric fluctuations, see, {\it
e.g.},~\cite{metric1}~\cite{metric2}.)  In order to study the production of
$\phi$
particles when the inflaton field is oscillating near the minimum of its
potential, we will need to calculate mode solutions for the quantum
fluctuations $\delta \phi$ before and after the phase transition,
matching the solutions by means of a Bogolyubov transformation. \\
\indent  We will first concentrate on the modes for the fluctuations
near the end of the inflationary period.  In terms of conformal time,
$d\eta \equiv a^{-1} dt$, the fluctuations are quantized as follows:
\beq
\delta \hat{\phi} (x) = \int \frac{d^3 \vec{k}}{(2 \pi)^{3/2}} \left[
\delta \phi_k (\eta) \> \hat{a}_{\vec{k}} \> e^{i \vec{k} \cdot \vec{x}} +
\delta \phi_k^* (\eta) \> \hat{a}^{\dagger}_{\vec{k}} \> e^{-i \vec{k} \cdot
\vec{x}} \right] ,
\label{qphi}
\eeq
where hats denote quantum operators, and the creation and annihilation
operators obey the canonical commutation relations:
\beqn
\nonumber \left[ \hat{a}_{\vec{k}} \> , \> \hat{a}_{\vec{\ell}} \right]
&=& \left[
\hat{a}^{\dagger}_{\vec{k}} \> , \> \hat{a}^{\dagger}_{\vec{\ell}} \right]
= 0 \>\> , \\
\left[ \hat{a}_{\vec{k}} \> , \> \hat{a}^{\dagger}_{\vec{\ell}} \right] &=&
\delta^3
(\vec{k} - \vec{\ell} ) \>\> , \>\> {\rm with} \>\> \hat{a}_{\vec{k}}
\vert 0 > = 0 .
\label{comrels}
\eeqn
{}From equation (\ref{eom}), and denoting $d/d\eta$ by a prime, the
equation of motion for the mode functions $\delta \phi_k (\eta)$ becomes:
\beq
\delta \phi_k^{\prime\prime} + 2 \frac{a^{\prime}}{a} \delta
\phi_k^{\prime} + \left( k^2 + 3 \lambda \varphi^2 a^2 \right) \delta
\phi_k = 0 .
\label{eom3}
\eeq
Next we introduce a \lq\lq conformal field" $\psi \equiv a \delta \phi$,
whose mode functions obey the equation of motion:
\beq
\psi_k^{\prime\prime} + \left[ k^2 - \frac{a^{\prime\prime}}{a} + 3
\lambda \varphi^2 a^2 \right] \psi_k = 0 .
\label{eom4}
\eeq
Near the end of inflation, the scale factor $a(t)$ will not in general
retain its de Sitter (exponential) form.  Instead we may write $a(t)
\propto t^p$, with $1/2 \leq p \leq \infty$, where $p = 1/2$ corresponds to
the
radiation-dominated epoch, and $p \rightarrow \infty$ recovers the de
Sitter epoch.  This scale factor corresponds to
a conformal-time scale factor of $a(\eta) \propto \eta^{p / (1 - p)}$.
Because we are ignoring the parametric resonance from the classical
$\varphi$-field's oscillations in this section, we may simply assume that
near the minimum of its potential, $\varphi^2 \ll k^2$.  If we lastly
define a new field $\chi \equiv \eta^{-1/2} \psi$, and define a new
variable $z \equiv k\eta$, then equation (\ref{eom4}) becomes
\beq
\frac{d^2 \chi_k}{dz^2} + \frac{1}{z} \frac{d \chi_k}{dz} + \left( 1 -
\frac{1}{z^2} \left[ \frac{(3p - 1)^2}{4 (p - 1)^2} \right] \right)
\chi_k \simeq 0 .
\label{eomBessel}
\eeq
This is now in the form of Bessel's equation.  Mode functions for the
original field $\delta \phi$ may then be written in terms of Hankel
functions:
\beqn
\nonumber \delta \phi_k (\eta) &=& \frac{\eta^{1/2}}{a (\eta)} \left[ A_k
H_{\nu}^{(1)} (k\eta) + B_k H_{\nu}^{(2)} (k\eta) \right] , \\
\nu &=& \frac{3p - 1}{2 (p - 1)} .
\label{phiBessel}
\eeqn
We assume that during inflation, the inflaton field is in its vacuum
state (see section 5.2 of~\cite{LidLyth} for further discussion of this
point).  The coefficients $A_k$ and $B_k$ thus may be fixed by
choosing an appropriate
quantum vacuum state.  The Bunch-Davies, or \lq\lq adiabatic", vacuum
requires that the fluctuations $\psi_k (\eta)$ behave as Minkowski-spacetime
mode functions far inside the horizon, that is, $\psi_k \rightarrow
(2k)^{1/2} \exp ( - ik\eta)$ for $k\eta \gg
1$.~\cite{metric2}~\cite{BirDavies}~\cite{BunchDavies}  From the asymptotic
properties of the Hankel functions, this gives
\beq
A_k = 0 \>\> , \>\> B_k = \frac{\sqrt{\pi}}{2} \exp \left[- i
\frac{\pi}{2} \left( \nu + \frac{1}{2} \right) \right] .
\label{AB}
\eeq
For the problem of (non-resonant) reheating, we will be interested in the
opposite asymptotic limit:  $k \eta \ll 1$, corresponding to
long-wavelength modes.  In this limit, the fluctuations $\psi_k (\eta)$ take
the form
\beq
\psi_k (\eta) \rightarrow \frac{2^{\nu - 1}}{\sqrt{\pi}} \exp \left[
-i \frac{\pi}{2} \left( \nu - \frac{1}{2} \right) \right] \Gamma (\nu)
\> \eta^{1/2} \> (k \eta)^{-\nu} .
\label{psilong1}
\eeq
In order to calculate the number ($N_k$) of $\phi$-particles produced per
mode $k$, we match this solution to long-wavelength mode solutions
following the phase transition into the radiation-dominated era. \\
\indent  To do this, we make use of a Bogolyubov transformation. (See,
{\it e.g.},~\cite{BirDavies}.)  For long-wavelength modes, we may
approximate the phase transition as instantaneous, occuring at some time
$\eta_*$~\cite{damvil}:
\beqn
\nonumber \eta < \eta_* &:& a (\eta) =  (a_o \eta )^{p / (1 - p)} \\
\eta > \eta_* &:& a(\eta) = C \left( \eta - \bar{\eta} \right) ,
\label{apt}
\eeqn
where $\bar{\eta} \equiv \eta_* - (a_o^2 \eta_*)^{-1}$.  It is convenient
to set $a(\eta_*) = 1$, which sets $\eta_* = a_o^{-1}$, and thus
$\bar{\eta} = 0$ and $C =
a_o$.  (Note that in the limit $p \rightarrow \infty$, the reference
scale $a_o$ becomes $-H_o$, where $H_o$ is the Hubble constant of
the de Sitter spacetime.)  Next we perform a Bogolyubov transformation to
match the long-wavelength mode solutions in the radiation-dominated era:
\beq
\psi_k (\eta > \eta_*) = \frac{1}{\sqrt{2k}} \left[ \alpha_k e^{-i k
\eta} + \beta_k e^{+i k \eta} \right] .
\label{deltaphikRD}
\eeq
The Bogolyubov coefficients $\alpha_k$ and $\beta_k$ may be
determined by requiring that both $\psi_k$ and $\psi_k^{\prime}$ be
continuous at $\eta = \eta_*$.  Then, using $N_k =
\left| \beta_k \right|^2$, this gives
\beq
N_k = \left| \beta_k \right|^2 = \frac{4^{\nu - 3/2}}{\pi} \Gamma^2
(\nu) \left( \nu - \frac{1}{2} \right)^2 \left( \frac{k}{a_o} \right)^{-2\nu
-1} .
\label{Nnores}
\eeq
Equation (\ref{Nnores}) is the main result of this section, and can now be
compared with the cases in which the effects of the inflaton's parametric
resonance are included.

\section{Particle creation from Parametric Resonance, with $H \simeq 0$}
\indent  We return to equation (\ref{eom4}) for the fluctuations $\psi_k
(\eta)$.  Rather than neglect the $3 \lambda \varphi^2 (\eta) a^2
(\eta)$ term in this section, however, we study the effects of this term as
the inflaton field $\varphi$ oscillates near the minimum of its
potential.  If the frequency of these oscillations is large enough, then we
may expand the fields $\varphi (\eta)$ and $\varphi^2 (\eta)$ in
conformal-time harmonics, analogous to the cosmic-time harmonic
decomposition adopted in~\cite{STB}.  Keeping the lowest term, we may
write
\beq
\varphi^2 (\eta) \simeq \overline{\varphi^2} \cos (\gamma \eta) ,
\label{varphi2}
\eeq
where $\overline{\varphi^2}$ is a slowly-decreasing amplitude.  This
quasi-periodic approximation for $\varphi^2 (\eta)$ may be compared with
the numerical results (for Minkowski-spacetime) calculated in~\cite{DB}.
For this section, we will assume that $\overline{\varphi^2} \simeq$
constant.  (As emphasized in~\cite{linde}~\cite{DB2}, the chief
contributor to the decrease of $\overline{\varphi^2}$ over time is
the back-reaction from created particles, and will be further discussed
in the Appendix.)  Substituting this ansatz into equation (\ref{eom4})
yields:
\beq
\psi_k^{\prime\prime}  + \omega_k^2 \left[ 1 + g a^2 \cos (\gamma \eta) -
\frac{a^{\prime\prime}}{k^2 a} \right] \psi_k = 0 ,
\label{eomPR1}
\eeq
where
\beq
\omega_k^2 \equiv k^2 \>\>, \>\> g \equiv \frac{3\lambda
\overline{\varphi^2}}{\omega_k^2} .
\label{wkg}
\eeq
Note that by working in terms of conformal time and the \lq\lq conformal
field" $\psi$, the frequency $\omega_k$ equals the (constant) comoving
wavenumber $k$, and does not redshift with increasing $a(\eta)$.  All of
the effects of expansion are included in the explicit factors of
$a(\eta)$ and $a^{\prime\prime} (\eta)$.  When expansion is neglected ({\it
i.e.}, when we set $a = 1$), equation (\ref{eomPR1}) reduces to the
Mathieu equation, the solutions of which are exponentially unstable when
$g \ll 1$ and $\gamma = 2 \omega_k + \epsilon$, with $\epsilon \ll
\omega_k$.  To see this, we may proceed as follows.  Our treatment
follows the methods outlined in~\cite{L2}, and substantially reproduces
the results from the alternative approach adopted in~\cite{STB}. \\
\indent  We introduce the trial solution:
\beq
\psi_k (\eta) = c(\eta) f \left[ a(\eta) \right] \cos \left[ \left(
\omega_k + \frac{1}{2}
\epsilon \right) \eta \right] + d(\eta) f \left[ a(\eta) \right] \sin \left[
\left( \omega_k + \frac{1}{2} \epsilon \right) \eta \right] .
\label{psiPRtrial}
\eeq
Here we have introduced the \lq scaling' function $f (a)$ to absorb effects
from the
expansion of the universe.  We assume that the coefficients $c(\eta)$ and
$d (\eta)$ are slowly-varying with time, as compared with the frequency
$\gamma / 2$, although we make no such assumptions about the behavior of
$f (a)$.  Then it is self-consistent to put $c^{\prime}, \> d^{\prime}
\sim {\it O} (\epsilon)$, and to neglect higher derivatives.  In
general, resonant solutions can be found for frequencies $n \gamma / 2
\simeq \omega_k$, but each integer $n \geq 1$ corresponds to keeping
terms of ${\it O} (g^n)$ in the perturbation expansion.~\cite{L2}  Since
we are only going to keep terms to ${\it O}(g)$, we may make the
following approximations:
\beqn
\nonumber \cos (\gamma \eta / 2) \cos (\gamma \eta) &=& \frac{1}{2} \cos
(\gamma \eta / 2) + \frac{1}{2} \cos (3 \gamma \eta / 2) \simeq
\frac{1}{2} \cos (\gamma \eta / 2) , \\
\sin (\gamma \eta / 2) \cos (\gamma \eta) &=& - \frac{1}{2} \sin (\gamma
\eta / 2) + \frac{1}{2} \sin (3 \gamma \eta / 2) \simeq - \frac{1}{2}
\sin (\gamma \eta / 2) .
\eeqn
Then the equation of motion (\ref{eomPR1}) applied to the trial solution
of equation (\ref{psiPRtrial}) yields:
\beq
\omega_k A (\eta)\> \sin (\gamma \eta / 2) + \omega_k
B(\eta)\> \cos (\gamma \eta / 2) + {\cal
O}\left(\epsilon^2 , \> g^2 \right) = 0 ,
\label{PReom2}
\eeq
with the coefficients $A$ and $B$ given by
\beqn
\nonumber A &=& -2 c^{\prime} f - c f^{\prime} \left( 2 +
\frac{\epsilon}{\omega_k} \right) + \frac{1}{\omega_k} \left( 2
d^{\prime} f^{\prime} + d f^{\prime\prime} \right)
- df \left( \frac{1}{2} g \omega_k a^2 + \epsilon +
\frac{1}{\omega_k} \frac{a^{\prime\prime}}{a} \right) , \\
B &=& 2 d^{\prime} f + df^{\prime} \left( 2 +
\frac{\epsilon}{\omega_k} \right) + \frac{1}{\omega_k} \left( 2
c^{\prime} f^{\prime} + c f^{\prime\prime} \right)
+ cf \left( \frac{1}{2} g \omega_k a^2 - \epsilon -
\frac{1}{\omega_k} \frac{a^{\prime\prime}}{a} \right) .
\label{AB2}
\eeqn
For equation (\ref{PReom2}) to be satisfied, we require that both of the
coefficients ($A$, $B$) of the trigonometric terms vanish identically.
This gives a pair of coupled differential equations involving the
coefficients $c (\eta)$, $d (\eta)$, and $f (a)$.  \\
\indent  For the remainder of this section we will neglect all expansion
of the universe, and set $f (a) = a = 1$.  Then, if we set $c (\eta) = C
\exp ( s \eta)$ and $d(\eta) = D \exp (s \eta)$, equations
(\ref{PReom2}) and (\ref{AB2}) yield
\beqn
\nonumber s C + \frac{1}{2} \left( \frac{1}{2} g \omega_k + \epsilon
\right) D &=& 0 , \\
s D + \frac{1}{2} \left( \frac{1}{2} g \omega_k - \epsilon \right) C
&=& 0 .
\label{cd}
\eeqn
These equations may be solved with
\beqn
\nonumber \frac{D}{C} &=& \mp \frac{1}{y_k} \equiv \mp \sqrt{\frac{
\frac{1}{2} g \omega_k - \epsilon}{ \frac{1}{2} g \omega_k + \epsilon}} , \\
s &=& \pm \frac{1}{2} \sqrt{ \left( \frac{1}{2} g \omega_k \right)^2 -
\epsilon^2 } .
\label{yso}
\eeqn
Thus there exists a growing solution and a decaying solution:
\beq
\psi_k^{\pm} (\eta) = \frac{\exp \left[ \pm s \eta
\right]}{\sqrt{\gamma y_k}} \left[ y_k \cos (\gamma \eta / 2) \mp \sin
(\gamma \eta / 2 ) \right] .
\label{psiPRnoexp}
\eeq
This normalization was chosen, following~\cite{TB90}, so that
$\psi_k^{+\>\prime} \psi_k^{-} - \psi_k^{-\>\prime} \psi_k^{+} = -1$. \\
\indent  From equations (\ref{yso}) and (\ref{psiPRnoexp}) it is clear
that the solutions $\psi_k$ will be exponentially unstable whenever $s$
is real.  Thus, the parametric resonance will occur only within a small
frequency band:
\beq
\left| \epsilon \right| < \frac{1}{2} g \omega_k .
\label{resband1}
\eeq
It will be convenient to introduce the variable $\ell$ as
\beq
\ell \equiv \frac{2 \epsilon}{g \omega_k} ,
\label{ell}
\eeq
so that resonance occurs when $-1 < \ell < 1$.  In terms of $\ell$, $s$
and $y_k$ may be rewritten
\beq
s =  \frac{g \omega_k}{4} \sqrt{1 - \ell^2} \>\>,\>\> y_k = \sqrt{
\frac{1 + \ell}{1 - \ell}} .
\label{ysoell}
\eeq
Note that $y_k \rightarrow 1$ as $\ell \rightarrow 0$, near the center of
the resonance band.  We further introduce the function $S_{\pm} (\eta)$ as
\beq
S_{\pm} (\eta) \equiv y_k \cos (\gamma \eta / 2) \mp \sin (\gamma \eta /
2) ,
\label{S}
\eeq
so that near the center of the resonance band, as $\ell \rightarrow 0$,
\beq
\left| S_{\pm} (\eta) \right|^2 \rightarrow 1 \mp
\sin (\gamma \eta) ,
\eeq
or, averaging over a few oscillations near $\ell \sim 0$,
\beq
\langle \left| S_{\pm} (\eta) \right|^2 \rangle \rightarrow 1 .
\eeq
Near the center of the resonance band, then, the solutions $\psi_k^{\pm}
(\eta)$ become
\beq
\psi_k^{\pm} (\eta) \rightarrow \frac{\exp \left[ \pm s \eta
\right]}{\sqrt{2k}} \> S_{\pm} (\eta) \simeq \frac{ \exp \left[ \pm s
\eta \right]}{\sqrt{2k}} ,
\eeq
where the irrelevant phase from the $S_{\pm}$ term has been dropped. \\
\indent  We can now make use of a Bogolyuobov transformation to solve for
the number of particles per mode, $N_k^{\rm res}$, produced by the
decaying inflaton field during its resonant oscillations.  In section 2,
we made the approximation of an instantaneous phase transition at some
time $\eta_*$, which can only be appropriate for long-wavelength modes.
In the present case, however, the resonant modes cannot have arbitrarily
long
wavelengths:  from equations (\ref{wkg}) and (\ref{resband1}), we require $k
\geq 2 \vert \epsilon \vert / g$.  Instead, use may be made of the more
elaborate time-dependent Bogolyubov transformation developed in Appendix
B of~\cite{STB}, the results of which are:
\beq
N_k = \left| \beta_k \right|^2 = \frac{1}{1 - \ell^2} \sinh^2 (s \eta ) .
\label{Npres1}
\eeq
As explained in~\cite{STB}, when $s$ is not exactly constant in time, but
changes adiabatically (such that $\vert s^{\prime} \vert \ll s^2$),
then this equation for $N_k$ may be modified to
\beq
N_k \simeq \sinh^2 \left( \int_{\rm res. \> band} s\> d\eta \right) ,
\label{Npres2}
\eeq
where the integral extends over the resonance band (equation
(\ref{resband1})).  Following~\cite{STB}, the (divergent) coefficient $(1 -
\ell^2)^{-1}$
has been dropped, because near $\vert \ell \vert \simeq 1$, the exponent
$s$ no longer changes adiabatically. \\
\indent  Equation (\ref{Npres2}), with $s$ given by equation
(\ref{yso}), is the main result of this section.  Clearly, if $\int s
d\eta$ is large enough within the resonance band, then the number of
particles produced will be exponentially greater than the number produced
from the non-resonant decay studied in section 2, given by equation
(\ref{Nnores}).  There is an important difference, however, between our
result for $N_k$ in equation (\ref{Npres2}) and the corresponding
expression in~\cite{STB}.  The analysis in~\cite{STB}
was carried out in terms of cosmic time $t$ and the fluctuations
$\delta \phi_k$; the equations of motion thus resemble equations
(\ref{eomH}) and (\ref{Mathieu}) above, with $\omega_k \simeq k/a$, and with
the assumption $\omega_k \gg H$.  As studied in~\cite{STB},
then, the resonant modes exit the resonance band due both to the redshift
of the physical wavenumbers $k/a$, as well as to the back-reaction
from produced particles.  This is to be
contrasted with the foregoing analysis:  In our case, by working in
terms of conformal time and the \lq\lq conformal field" $\psi_k$, the
only time-dependence of the exponent $s$ comes from the slow
time-dependence of the decaying amplitude $\overline{\varphi^2}$.  That
is, in our case it is $g$ which is slowly changing with time, and not
$\omega_k$; and it is this change of $g$ which causes various modes
$\psi_k^+$ to slide outside of the resonance band. The time-dependence of
$g$ is further addressed in the Appendix.  Meanwhile, we will assume for
the remainder of this paper that $\overline{\varphi^2} (\eta)$, and hence
$g (\eta)$ and $s (\eta)$, may be treated as slowly-varying with time. With
this assumption, we
may now study what happens to the resonant modes when the expansion of the
universe is included.

\section{Resonant Inflaton Decay in an Expanding Background}
\indent We return to the full equation of motion for the fluctuations
$\psi_k$, equation (\ref{eomPR1}), with $\omega_k$ and $g$ as defined in
equation (\ref{wkg}).  We introduce the same trial solution, equation
(\ref{psiPRtrial}), and keep terms to ${\it O}(\epsilon , \> g)$.  This
leads to equation (\ref{PReom2}), with the full coefficients $A(\eta)$
and $B (\eta)$ given by equation (\ref{AB2}).  We saw in the previous
section that a consistent resonant solution may be found when $f (a)$ and
$a$ are both set equal to unity.  In this section, we keep these terms
explicit and general, and show that consistent resonant solutions may be
found when the expansion of the universe is included. \\
\indent  We split the coefficients $A$ and $B$ into two terms each: $A =
A_1 + A_2$, and $B = B_1 + B_2$, with
\beqn
\nonumber A_1 &=& -2 c^{\prime} f - df \left( \frac{1}{2} g \omega_k +
\epsilon \right) , \\
\nonumber A_2 &=& - c f^{\prime} \left( 2 + \frac{\epsilon}{\omega_k}
\right) +
\frac{1}{\omega_k} \left( 2 d^{\prime} f^{\prime} + d f^{\prime\prime}
\right) - df \left( \frac{1}{2} g \omega_k \left( a^2 - 1 \right) +
\frac{1}{\omega_k} \frac{a^{\prime\prime}}{a} \right) , \\
\nonumber B_1 &=& 2 d^{\prime} f + cf \left( \frac{1}{2} g \omega_k -
\epsilon \right) , \\
B_2 &=& d f^{\prime} \left( 2 + \frac{\epsilon}{\omega_k} \right) +
\frac{1}{\omega_k} \left( 2 c^{\prime} f^{\prime} + c f^{\prime\prime}
\right) + cf \left( \frac{1}{2} g \omega_k \left( a^2 - 1 \right) -
\frac{1}{\omega_k} \frac{a^{\prime\prime}}{a} \right) .
\label{AB3}
\eeqn
In order for equation (\ref{PReom2}) to be satisfied, we again require
that the coefficients vanish:  $A = B = 0$.  Because we are interested in
resonant behavior, we will further look for solutions when we set each of
the terms $A_1$, $A_2$, $B_1$, and $B_2$ separately equal to zero.  Then the
coefficients $A_1$ and $B_1$ reduce to the same pair of linear
differential equations as in equation (\ref{cd}), which may be solved with
the same $y_k$ and $s$ as given in equation (\ref{yso}).  The task now is
to use the equations
\beqn
\nonumber  - c \left( \frac{f^{\prime}}{f} \right) \left( 2 +
\frac{\epsilon}{\omega_k} \right) +
\frac{1}{\omega_k} \left[ 2 d^{\prime} \left( \frac{f^{\prime}}{f} \right)
+ d \left( \frac{f^{\prime\prime}}{f} \right)
\right] - d \left( \frac{1}{2} g \omega_k \left( a^2 - 1 \right) +
\frac{1}{\omega_k} \frac{a^{\prime\prime}}{a} \right)  &=& 0 , \\
d \left( \frac{f^{\prime}}{f} \right) \left( 2 +
\frac{\epsilon}{\omega_k} \right) + \frac{1}{\omega_k} \left[ 2
c^{\prime} \left( \frac{f^{\prime}}{f} \right) + c \left(
\frac{f^{\prime\prime}}{f} \right) \right] + c \left( \frac{1}{2} g
\omega_k \left( a^2 - 1 \right) -
\frac{1}{\omega_k}\frac{a^{\prime\prime}}{a} \right) &=& 0 ,
\label{A2B2}
\eeqn
to find a solution for the expansion factor, $f [a (\eta)]$. \\
\indent  The first step is to notice that several terms in equation
(\ref{A2B2}) may be combined:
\beq
\left[ 2 d^{\prime} \left( \frac{f^{\prime}}{f} \right) + d \left(
\frac{f^{\prime\prime}}{f} \right) \right] = \frac{1}{\left(fd\right)}
\frac{d}{d\eta} \left( d^2 f^{\prime} \right) ,
\eeq
and similarly for the corresponding $cf$-term in the second line of
equation (\ref{A2B2}).  Then, because the coefficients $c(\eta)$ and
$d(\eta)$ are determined by equation (\ref{cd}), the two equations in
(\ref{A2B2}) may be combined to give
\beqn
\nonumber \frac{d}{d\eta} \left[ e^{2s\eta} f^{\prime} \right] &=&
\left[ \frac{a^{\prime\prime}}{a} + \frac{1}{2} \frac{ \left( y_k^2 - 1
\right)}{\left( y_k^2 + 1 \right)} g \omega_k^2 \left( 1 - a^2 (\eta)
\right) \right] e^{2 s \eta} f(\eta) \\
\nonumber &=& \left[ \alpha (p) \> \eta^{-2} + \frac{1}{2} g \ell \omega_k^2
\left( 1 - a^2 (\eta) \right) \right] e^{2 s \eta} f (\eta ) \\
&\equiv& E^2 (\eta) e^{2s\eta} f (\eta) .
\label{df1}
\eeqn
In the second line, we have used the definitions of $y_k$ in equation
(\ref{yso}) and $\ell$ in equation (\ref{ell}), and have introduced the
coefficient
$\alpha (p)$.  For a cosmic-time scale factor $a (t) \propto t^p$, or,
equivalently, a conformal-time scale factor $a
(\eta) = (a_o \eta)^{p / (1 - p)}$, the coefficient $\alpha (p)$ is defined
as:
\beq
\frac{a^{\prime\prime}}{a} = \frac{p (2p - 1)}{(p - 1)^2}\> \eta^{-2}
\equiv \alpha (p) \>\eta^{-2} .
\label{alpha}
\eeq
Note that $\alpha (p) \geq 0$ for $p \geq 1/2$, and that for de Sitter
expansion, $\alpha (\infty) = 2$. \\
\indent  From equation (\ref{df1}), we may write the second-order
differential equation for $f$:
\beq
f^{\prime\prime} + 2 s f^{\prime} - E^2 (\eta) f = 0 .
\label{d2f}
\eeq
If we now use the fact that $s \sim {\it O}(g, \> \epsilon)$, and that
$g \ll 1$, then we may approximate\footnote{This approximation for $E^2
(\eta)$ is further justified far from the edge of the resonance band,
when $\left| \ell \right| \ll 1$.  Also, we may assume,
as in section 2, that the phase transition occurs near $a(\eta_*) = 1$.}
 $E^2 (\eta) \simeq \alpha (p) \>\eta^{-2}$, and
\beq
f^{\prime\prime} -  \alpha (p) \>\eta^{-2}f \simeq 0 .
\label{d2f2}
\eeq
This equation may be solved with the ansatz $f (\eta) = \beta \>
\eta^{\mu_{1,\>2}}$, with
\beq
\mu_{1,\> 2} = \frac{1}{2} \left[ 1 \pm \sqrt{1 + 4
\alpha (p)} \right] .
\label{mu1}
\eeq
The two possible solutions for $\mu$, labeled by subscript 1 and 2,
correspond to the choice of $\pm$ in equation (\ref{mu1}); the choice
of whether $\mu_1$ or $\mu_2$ is appropriate will be determined below.  With
this
solution for $f (\eta)$, the resonant solutions $\psi_k$ may be written:
\beq
\psi_k^{\pm} (\eta) = \frac{f (\eta) \exp [ \pm s\eta]}{\sqrt{\gamma
y_k}} S_{\pm} (\eta) ,
\eeq
with $S_{\pm} (\eta)$ given by equation (\ref{S}). \\
\indent  We may now choose the
appropriate exponent $\mu_{1,\>2}$ for the function $f$ by matching the
$\eta$-dependence of this resonant solution for $\psi_k^{\pm}$ with the
non-resonant solutions found in section 2.  For fixed $\eta$, the
(non-resonant) long-wavelength solutions behave as (see equations
(\ref{phiBessel}) and (\ref{psilong1})):
\beq
\psi_k^{\rm nr} \propto \eta^{p / (1 - p)} .
\label{psinr}
\eeq
In the resonant case considered in this section,
$\psi_k^{\pm} \propto f \propto \eta^{\mu_{1,\>2}}$.  From the definition
of $\alpha (p)$ in equation (\ref{alpha}), the two exponents may be written
\beq
\mu_1 = \frac{1 - 2p}{1 - p} \>\>,\>\> \mu_2 =
\frac{p}{1 - p} .
\eeq
Thus the appropriate exponent for the resonant case is $\mu_2$.  Note
that $f (\eta)$ has the
same $\eta$-dependence as the conformal-time scale factor, $a (\eta)$.
Because we require that $f [a(\eta)] = 1$ when $a (\eta) = 1$, we may fix
the normalization of $f$ as $f = a (\eta) = (a_o \eta)^{\mu_2}$.
In addition, given that $\alpha (p) \geq 0$ for $p \geq 1/2$, we will
write this exponent as
\beq
\mu_2 = - \left| \mu_2 \right| = - \frac{1}{2} \left[ \sqrt{ 1 + 4 \alpha
(p)} - 1 \right], \label{mu2}
\eeq
and thus $f = (a_o \eta)^{- \left| \mu_2 \right|}$.\\
\indent  With the \lq scaling' function $f$ now determined,
the full resonant solutions in an expanding background may be written:
\beq
\psi_k^{\pm} (\eta) = \frac{\exp \left[ \pm x_{\pm} \eta
\right]}{\sqrt{\gamma y_k}} S_{\pm} (\eta) ,
\label{psiexp}
\eeq
with the exponent $x_{\pm} (\eta)$ defined by:
\beq
x_{\pm} (\eta) \equiv \pm s - \frac{\left|\mu_2\right|}{\eta} \ln (a_o
\eta) .
\label{x}
\eeq
The $s$-term is given by equation (\ref{yso}).  As in the previous
section, resonance occurs when $2 \vert \epsilon \vert < g \omega_k$.
With the fluctuations $\psi_k^{\pm} (\eta)$ now in the form of equation
(\ref{psiexp}), we may again use the result of Appendix B of~\cite{STB},
and solve for the number of $\phi$-particles produced per mode $k$:
\beq
N_k \simeq \sinh^2 \left( \int_{\rm res. \> band} x_+ (\eta) \> d\eta
\right) .
\label{Nprexp}
\eeq
{}From the form of $\mu_2$
in equation (\ref{mu2}), it is clear that the expansion of the universe
decreases the exponential production:  $\int x_+ d\eta$ over the resonance
band is less than $\int s d\eta$, corresponding to the non-expanding case
of section 3.
Nevertheless, it is possible analytically to find self-consistent resonant
solutions for
the fluctuations $\psi_k^{\pm}$ when the expansion of the universe is
included.

\section{Conclusion}
\indent  We have demonstrated that the resonance effects studied
in~\cite{linde}~\cite{STB}~\cite{DB} can indeed have dramatic
consequences for post-inflationary reheating.  In particular, it has been
shown that solutions for the number of particles produced from
the resonant decay of the inflaton field may be found given a general
background expansion, $a(t) \propto t^p$.  Although the total number of
particles produced in this
expanding case, $N_k$ in equation (\ref{Nprexp}), is less than the
corresponding expression when $H \simeq 0$ (equation \ref{Npres2}),
there still remain regions of parameter space for which
this resonant production far outweighs the non-resonant production
described in section 2. The potential trouble for the new resonant
reheating scenario is therefore not expansion of the universe, but rather
the decaying amplitude of the classical $\varphi$-field.  This question
is discussed further below, in the Appendix.\\
\indent  As pointed out in~\cite{linde}, the case of inflaton decay into
further inflatons (as studied above) may be of interest for dark matter
searches.  Following their production, the inflatons would decouple from
the rest of matter.  If the
inflatons were given a tiny mass (which has been neglected here), then
these bosons could serve as a natural candidate for the missing dark matter.
Furthermore, much of the formalism developed in this paper may be taken
over,
unchanged, for studying the case of inflaton decay into some distinct
light species of boson, as studied in~\cite{linde}~\cite{STB}.  In such
decays, it is
again expected that the parametric resonance would survive a non-zero
expansion of the universe.  The difference for these decays would reside
solely in the form of $\overline{\varphi^2} (\eta)$. \\
\indent  The foregoing analysis for the simple
model of a minimally-coupled scalar field $\phi$ with an Einstein-Hilbert
gravitational action can also be applied directly to many classes of
Generalized Einstein Theory inflationary models, such as those studied
in~\cite{DKprd2}.  For these models, the conformal transformation factor
$\Omega (x) \rightarrow 1$ as the inflaton field reaches the minimum of
its potential.  During the $\phi$-field's oscillatory phase,
the non-minimal $\phi^2 R$ coupling takes
the form of the standard $(16\pi G)^{-1}R$ gravitational action of a
minimally-coupled theory.  Near the epoch of reheating, then, the
effective lagrangian takes the form of equation (\ref{lag}), from which
the reheating analysis may proceed as above.  It would be interesting to
calculate
$N_k$ for the new models of open inflation~\cite{openi}, following the
methods developed above, which might display qualitatively different
reheating scenarios.  This is the subject of further study.

\section{Appendix}
\indent  It is
important to understand how $\overline{\varphi^2}$, and hence $g$ and $s$,
changes with time.  A proper treatment would be to extend the
Minkowski-spacetime approach adopted in~\cite{DB}~\cite{DB2}, and to work
out the
thermal Green's functions for the decaying inflaton field for the general
background expansion considered here.  (See also~\cite{phaset}.)  As a
first approximation, we may posit the reasonable phenomenological ansatz,
which may be compared with figure 1 in~\cite{DB}:
\beq
\overline{\varphi^2} (\eta) \simeq \varphi_o^2 \> e^{-\eta / \tau} ,
\label{exp}
\eeq
where $\tau$ is some damping time-scale, most likely to be determined
from numerical integration.  In other words, we will assume here that
the frequency of the $\cos (\gamma \eta )$ oscillations may be treated
as nearly constant, while the amplitude of the oscillations decreases.  From
figure 1 in~\cite{DB}, it is clear
that we may assume that $\tau^{-1} \ll \omega_k$.  If we then write $g
(\eta) = g_o \exp ( - \eta / \tau)$, and keep terms to first order in $g_o$,
$(\epsilon / \omega_k)$, and $( \tau^{-1} / \omega_k )$, we may repeat the
above analysis
in equations (\ref{eomPR1})-(\ref{psiPRnoexp}).  We will restrict
attention to the non-expanding case, since the expanding case examined in
section 4 reveals the same dependence of $N_k$ on $g (\eta)$. \\
\indent  We introduce the trial solution
\beq
\psi_k (\eta) = c(\eta) e^{-\eta / \tau} \cos (\gamma \eta / 2) + d(\eta)
e^{-\eta / \tau} \sin (\gamma \eta / 2) .
\label{psiapp1}
\eeq
Again taking $c^{\prime}, \> d^{\prime} \sim {\it O} (\epsilon)$
leads to the coupled differential equations:
\beqn
\nonumber -2 c^{\prime} + 2 c \tau^{-1} - d \epsilon - \frac{1}{2} d g_o
\omega_k e^{-\eta / \tau} &=& 0 , \\
2 d^{\prime} - 2 d \tau^{-1} - c \epsilon + \frac{1}{2} c g_o \omega_k
e^{-\eta / \tau} &=& 0 .
\label{apdifeq}
\eeqn
These equations may be solved approximately by writing $c (\eta) = C \exp [
\theta \eta ]$ and $d (\eta) = D \exp [ \theta \eta]$, with
\beqn
\nonumber \frac{D}{C} = \mp \frac{1}{y_k (\eta)} \equiv \mp \sqrt{
\frac{\frac{1}{2} g_o
\omega_k \exp [- \eta / \tau] - \epsilon}{\frac{1}{2} g_o \omega_k \exp
[- \eta / \tau] + \epsilon}} , \\
\theta (\eta) = \pm \frac{1}{2} \sqrt{ \left( \frac{1}{2} g_o \omega_k
\right)^2 e^{ - 2 \eta / \tau } - \epsilon^2} + \frac{1}{\tau} \equiv \pm
s (\eta) + \frac{1}{\tau} .
\label{ysoap}
\eeqn
With these solutions, the growing and decaying mode functions take the form:
\beq
\psi_k^{\pm} (\eta) = \frac{\exp [ \pm s (\eta) \> \eta ]}{\sqrt{\gamma y_k
(\eta)}}  \left[ y_k (\eta) \cos (\gamma
\eta / 2) \mp \sin (\gamma \eta / 2) \right] .
\label{psiap2}
\eeq
These solutions have the benefit over the corresponding solutions in
equation (\ref{psiPRnoexp}) of ending the resonance in a natural way:
unlike the
solutions from section 3, these mode functions do not continue to grow
exponentially forever.  Now the resonance band is explicitly time-dependent:
\beq
\left| \epsilon \right| < \frac{g_o \omega_k}{2} e^{-\eta / \tau} .
\label{resbandap}
\eeq
In addition, the slight time-dependence of $y_k
(\eta)$ and $s (\eta)$ still allows these mode functions to be normalized
as:
$\psi_k^{+ \> \prime} \psi_k^{-} - \psi_k^{-\> \prime} \psi_k^{+} = -1 +
{\it O} \left( \tau^{-1} / \omega_k \right)$. \\
\indent  With this simple ansatz for the decaying inflaton field
amplitude, the exponent $s (\eta)$ changes in time as
\beq
\frac{ \left| s^{\prime} \right|}{s^2} = \frac{ \tau^{-1}}{s (\eta)}
\frac{1}{ \left[ 1 - \ell^2 (\eta) \right]} .
\label{sprime}
\eeq
In analogy with the situation in section 3, we have defined the variable
$\ell (\eta) \equiv (2 \epsilon ) / (g (\eta) \omega_k)$.  In order to
satisfy the adiabaticity requirement for $s (\eta)$ (away from the edges
of the resonance band), we must require $\tau^{-1} < g_o \omega_k$.  Note
that this is more stringent than the original assumption, $\tau^{-1} \ll
\omega_k$.  If this new constraint may be satisfied, then the number of
$\phi$-particles produced may be calculated from:
\beqn
\nonumber \int_{\rm res.\> band} s (\eta) \> d \eta &=&  \frac{ \tau
\epsilon}{2} \int_{-1}^{+1} d\ell \frac{ \sqrt{1 - \ell^2}}{\ell^2} \\
&=& \frac{\tau \epsilon}{2} \left[ - \frac{ \sqrt{ 1 - \ell^2}}{\ell} -
\arcsin \left( \ell \right) \right]_{\ell = -1}^{\ell = +1} \\
&=& - \frac{ \pi \tau \epsilon}{2} .
\label{intrb}
\eeqn
This gives $N_k \simeq \sinh^2 ( \int s d\eta ) = \sinh^2 ( \pi \tau
\epsilon / 2)$.  Given the constraints on both $\tau$ and $\epsilon$, we
find $N_k \gg 1$.  \\
\indent  As emphasized in~\cite{linde}~\cite{DB2}, the decay of the
amplitude of the inflaton might instead take the form of a power-law in
time, rather than the exponential form adopted in equation (\ref{exp}).
If we write
\beq
\overline{\varphi^2} (\eta) \simeq \varphi_o^2 \left( \frac{\eta}{\eta_o}
\right)^{-b} ,
\label{powerlaw}
\eeq
with $b > 0$ a real constant, and thus
\beq
g (\eta) = g_o \left( \frac{\eta}{\eta_o} \right)^{-b} \>\> , \>\>
\ell (\eta) = \frac{2 \epsilon}{g_o \omega_k} \left( \frac{\eta}{\eta_o}
\right)^{b} ,
\label{gell}
\eeq
then we may again repeat the steps in equations
(\ref{eomPR1})-(\ref{psiPRnoexp}).  The solutions $y_k$ and $s$ again take
the form
\beq
y_k (\eta) = \sqrt{ \frac{ \frac{1}{2} g(\eta) \omega_k +
\epsilon}{ \frac{1}{2} g(\eta) \omega_k - \epsilon}} \>\>,\>\> s(\eta) =
\frac{1}{2} \sqrt{ \left( \frac{1}{2} g(\eta) \omega_k \right)^2 -
\epsilon^2} ,
\label{ys3}
\eeq
and the growing and decaying mode functions again take the form of
equation (\ref{psiap2}), with $y_k$ and $s$ now given by equation
(\ref{ys3}).  Now the exponent $s (\eta)$ changes in time as
\beq
\frac{ \left| s^{\prime} \right|}{s^2} = \frac{ b \eta^{-1}}{s
(\eta)} \frac{1}{ \left[ 1 - \ell^2 (\eta) \right]} .
\eeq
In order to satisfy the adiabaticity condition, we therefore require
$b < g_o$.  Then the number of $\phi$-particles produced may be
calculated from
\beqn
\nonumber \int_{\rm res.\> band} s (\eta) \> d\eta &=& \frac{\epsilon
\eta_o}{2 b} \left( \frac{g_o \omega_k}{2 \epsilon}
\right)^{1/ b} \int_{-1}^{+1} d \ell \> \ell^{1/ b} \> \frac{
\sqrt{1 - \ell^2}}{\ell^2} \\
&\simeq& {\it O} (1) \cdot \frac{\epsilon \eta_o}{2 b}
\> \ell_o^{-1/ b} ,
\eeqn
where $\ell_o \equiv \ell (\eta_o)$.  From equation (\ref{gell}), we see
that $\ell_o < \ell_{\rm max} = 1$, so that given $0 < b < g_o \ll
1$, $\ell_o^{-1/ b} \gg 1$.  This yields $N_k \sim \sinh^2 (\epsilon
\eta_o \ell_o^{-1/ b} / (2 b)) \gg 1$. \\
\indent  These two different forms for the decay of the inflaton field
due to back-reaction from produced particles, the exponential decay of
equation (\ref{exp}), and the power-law decay of equation
(\ref{powerlaw}), both allow a large production of particles from the
inflaton's decay, so long as the adiabaticity requirement on $s (\eta)$
may be met.  It is clear, however, that a more extended treatment
of the effects of back-reaction on the
decaying amplitude $\overline{\varphi^2} (\eta)$ is required, and is the
subject of further study.

\section{Acknowledgments}
\indent  This research was conducted at Dartmouth College.  It is a
pleasure to thank Joseph Harris, Marcelo Gleiser, and John Walsh for their
hospitality.  I would also like to thank Dan Boyanovsky for providing
copies of the figures which accompany ref.~\cite{DB}, and A. Linde,
L. Kofman, and A. Starobinsky for their helpful comments on an earlier
draft of this paper.
This work was supported by an NSF Fellowship for Pre-Doctoral Fellows,
and in part by NSF grant PHY-9218167.

%

%
%

\begin{thebibliography}{9999}
\baselineskip 14pt
\bibitem[1]{alan} A. H. Guth, \it Phys. Rev. D \bf 23\rm ,
347 (1981).
%
\bibitem[2]{newi} A. Linde, \it Phys. Lett. B \bf 108\rm , 389 (1982); \\
A. Albrecht and P. Steinhardt, \it Phys. Rev. Lett. \bf 48\rm , 1220 (1982) .
%
\bibitem[3]{reh} A. D. Dolgov and A. D. Linde, \it Phys. Lett. B \bf
116\rm , 329 (1982); \\
L. F. Abbott, E. Farhi, and M. B. Wise, \it Phys. Lett. B \bf 117\rm , 29
(1982).
%
\bibitem[4]{chaotic} A. D. Linde, \it Phys. Lett. B \bf 129\rm , 177
(1983).
%
\bibitem[5]{linde} L. Kofman, A. Linde, and A. A.
Starobinsky, \it Phys. Rev. Lett. \bf 73\rm , 3195 (1994).
%
\bibitem[6]{STB} Y. Shtanov, J. Traschen, and R.
Brandenberger, \it Phys. Rev. D. \bf 51\rm , 5438 (1995).
%
\bibitem[7]{DB} D. Boyanovsky, M. D'Attanasio, H. J. de Vega, R. Holman,
D.-S. Lee, and A. Singh, \lq\lq Reheating the Post Inflationary
Universe," Preprint hep-ph/9505220.
%
\bibitem[8]{DB2} D. Boyanovsky, M. D'Attanasio, H. J. de Vega, R.
Holman, and D.-S. Lee, \lq\lq Reheating and Thermalization:  Linear vs.
Non-Linear Relaxation,"  Preprint hep-ph/9507414.
%
\bibitem[9]{dolfre} A. Dolgov and K. Freese, \it Phys. Rev. D \bf 51\rm
, 2693 (1995).
%
\bibitem[10]{Yosh} M. Yoshimura, \lq\lq Catastrophic Particle Production
under Periodic Perturbation," Preprint TU/95/484, hep-th/9506176; \\
H. Fujisaki, K. Kumekawa, M. Yamaguchi, and M. Yoshimura, \lq\lq Particle
Production and Dissipative Cosmic Field," Preprint TU/95/488,
hep-ph/9508378.
%
\bibitem[11]{TB90} J. H. Traschen and R. H. Brandenberger,
\it Phys. Rev. D. \bf 42\rm , 2491 (1990).
%
\bibitem[12]{dolkir} A. D. Dolgov and D. P. Kirilova, \it Sov. J. Nucl.
Phys. \bf 51\rm , 172 (1990).
%
\bibitem[13]{handb} M. Abramowitz and I. Stegun, {\it Handbook of
Mathematical Functions} \rm (Dover, New York, 1972), pp. 722-746.
%
\bibitem[14]{L2} L. D. Landau and E. M. Lifshitz, {\it Mechanics} \rm
(Pergamon, New York, 1960), pp. 80-83.
%
\bibitem[15]{preheat} L. Kofman, A. Linde, and A. A. Starobinsky, \lq\lq
Non-Thermal Phase Transitions after Inflation," Preprint SU-ITP-95-21,
hep-th/9510119; \\
I. I. Tkachev, \lq\lq Phase Transitions at Preheating," Preprint
OSU-TA-21/95, hep-th/9510146.
%
\bibitem[16]{metric1} V. F. Mukhanov, \it JETP Lett. \bf 41\rm , 493
(1985); \it idem.\rm , \it Phys. Lett. B \bf 218\rm , 17 (1989); \\
V. F. Mukhanov, H. A. Feldman, and R. H. Brandenberger, \it Phys. Rep.
\bf 215\rm , 203 (1992).
%
\bibitem[17]{metric2} M. Sasaki, \it Prog. Theo. Phys. \bf 76\rm , 1036
(1986); \\
J.-C. Hwang, \it Phys. Rev. D \bf 48\rm , 3544 (1993); \it idem.\rm , \it
Ap. J. \bf 427\rm , 542 (1994).
%
\bibitem[18]{BirDavies} N. D. Birrell and P. C. W. Davies, {\it Quantum
Fields in Curved Space} (Cambridge University Press, 1982).
%
\bibitem[19]{LidLyth} A. R. Liddle and D. H. Lyth, \it Phys. Rep. \bf
231\rm , 1 (1993).
%
\bibitem[20]{BunchDavies} T. S. Bunch and P. C. W. Davies, \it Proc. Roy.
Soc. A \bf 360\rm , 117 (1978); \it idem.\rm , \it J. Phys. A \bf 11\rm ,
1315 (1978); \\
See also R. H. Brandenberger, \it Nucl. Phys. B \bf 245\rm , 328 (1984); \\
A. H. Guth and S.-Y. Pi, \it Phys. Rev. D \bf 32\rm , 1899 (1985).
%
\bibitem[21]{damvil} T. Damour and A. Vilenkin, \lq\lq String Theory and
Inflation," Preprint hep-th/9503149.
%
\bibitem[22]{DKprd2} D. I. Kaiser, \it Phys. Rev. D \bf 52\rm , 4295
(1995).
%
\bibitem[23]{openi} See, {\it e.g.}, M. Bucher, A. S. Goldhaber, and N.
Turok, \lq\lq An Open Universe from Inflation," Preprint
iassns-hep-94-81, PUPT-94-1507, hep-ph/9411206; \\
A. Linde, \it Phys. Lett. B \bf 351\rm , 99 (1995); \\
A. Linde and A. Mezhlumian, \lq\lq Inflation with $\Omega \neq 1$,"
Preprint SU-ITP-95-11, astro-ph/9506017.
%
\bibitem[24]{phaset} D. Boyanovsky and H. J. de Vega, \it Phys. Rev. D
\bf 47\rm , 2343 (1993); \\
D. Boyanovsky, D.-S. Lee, and A. Singh, \it Phys. Rev. D \bf 48\rm , 800
(1993); \\
D. Boyanovsky, H. J. de Vega, and R. Holman, \it Phys. Rev. D \bf 49\rm ,
2769 (1994); \\
D. Boyanovsky, H. J. de Vega, R. Holman, D.-S. Lee, and A. Singh, \it
Phys. Rev. D \bf 51\rm , 4419 (1995); \\
See also Marcelo Gleiser, \lq\lq Thermal Mixing of Phases:  Numerical and
Analytical Studies," Preprint DART-HEP-95/03, hep-ph/9507312, and
references therein.
%

%
\end{thebibliography}
\end{document}